\documentclass[conference]{IEEEtran}
\IEEEoverridecommandlockouts

\usepackage{float}
\usepackage{booktabs}
\usepackage{cite}
\usepackage{amsmath,amssymb,amsfonts}
\usepackage{algorithmic}
\usepackage{graphicx}
\usepackage{textcomp}
\usepackage{xcolor}
\usepackage{makecell}
\usepackage{caption}
\usepackage{stfloats}
\usepackage{enumitem}
\usepackage{array} 
\def\BibTeX{{\rm B\kern-.05em{\sc i\kern-.025em b}\kern-.08em
    T\kern-.1667em\lower.7ex\hbox{E}\kern-.125emX}}

\begin{document}

\title{LoVA: Long-form Video-to-Audio Generation}
\author{
    \IEEEauthorblockN{
        Xin Cheng, 
        Xihua Wang, 
        Yihan Wu, 
        Yuyue Wang and
        Ruihua Song\IEEEauthorrefmark{2}
    }\\
    \vspace{-0.8em}
    \IEEEauthorblockA{
        Gaoling School of Artificial Intelligence, Renmin University of China, Beijing 100872, China\\
        Email: 
            \{chengxin000, xihuaw, yihanwu, wangyuyue123, rsong\}@ruc.edu.cn
    }

    \thanks{
    \vspace{1mm}
    \hspace{-\dimexpr\oddsidemargin+0.5in} 
    \rule{0.5\textwidth}{0.4pt} 

    \IEEEauthorrefmark{2}Corresponding author.
    }
    
    \vspace{-1em}

}
\maketitle

\begin{abstract}

Video-to-audio (V2A) generation is important for video editing and post-processing, enabling the creation of semantics-aligned audio for silent video. However, most existing methods focus on generating short-form audio for short video segment (less than 10 seconds), while giving little attention to the scenario of long-form video inputs. For current UNet-based diffusion V2A models, an inevitable problem when handling long-form audio generation is the inconsistencies within the final concatenated audio. In this paper, we first highlight the importance of long-form V2A problem. Besides, we propose LoVA, a novel model for \textbf{Lo}ng-form \textbf{V}ideo-to-\textbf{A}udio generation. Based on the Diffusion Transformer (DiT) architecture, LoVA proves to be more effective at generating long-form audio compared to existing autoregressive models and UNet-based diffusion models.
Extensive objective and subjective experiments demonstrate that LoVA achieves comparable performance on 10-second V2A benchmark and outperforms all other baselines on a benchmark with long-form video input. 

\end{abstract}


\section{Introduction}

Video-to-Audio (V2A) generation, which aims to create synchronized and realistic sound effects for silent videos, finds widespread use in video editing, sound effect creation, and autonomous content enhancement~\cite{ament2014foley}. 
However, current V2A methods predominantly focus on generating fixed-length short audios, typically less than 10 seconds on VGGSound~\cite{chen2020vggsound} or Audioset~\cite{gemmeke2017audio} benchmarks. 
These methods generate fixed-length audios through autoregressive approaches truncated to a maximum length~\cite{SpecVQGAN_Iashin_2021,mei2023foleygen,sheffer2023hear}, 
or through UNet-based diffusions to denoise fixed-length noise~\cite{luo2024diff,zhang2024foleycrafter,wang2024tiva}.
Despite their success in generating fixed-length short audios, the challenge of creating audio for variable-length, long-form videos exceeding 10 seconds in real-world scenarios remains unexplored.
Our work aims to address this short-to-long duration gap in the V2A domain.

\begin{figure}[htbp]
\centerline{\includegraphics[width=1\linewidth]{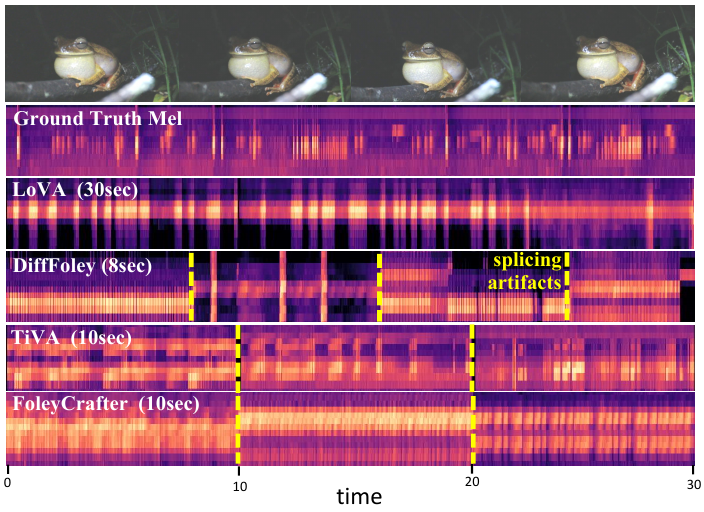}}
\vspace{-0.2cm}
\caption{\fontsize{9pt}{11pt} \selectfont 
Long-form V2A example. Current (8s/10s) UNet-based diffusion V2A models (DiffFoley, TiVA, FoleyCrafter) exhibit inconsistency when generating long-form (30s) audio, as indicated by clear mel-spectrogram boundaries and structural variances. In contrast, our LoVA produces consistent results similar to the ground truth.
}
\vspace{-0.5cm}
\label{case}
\end{figure}

\begin{figure*}[ht]
\centerline{\includegraphics[width=0.95 \linewidth]{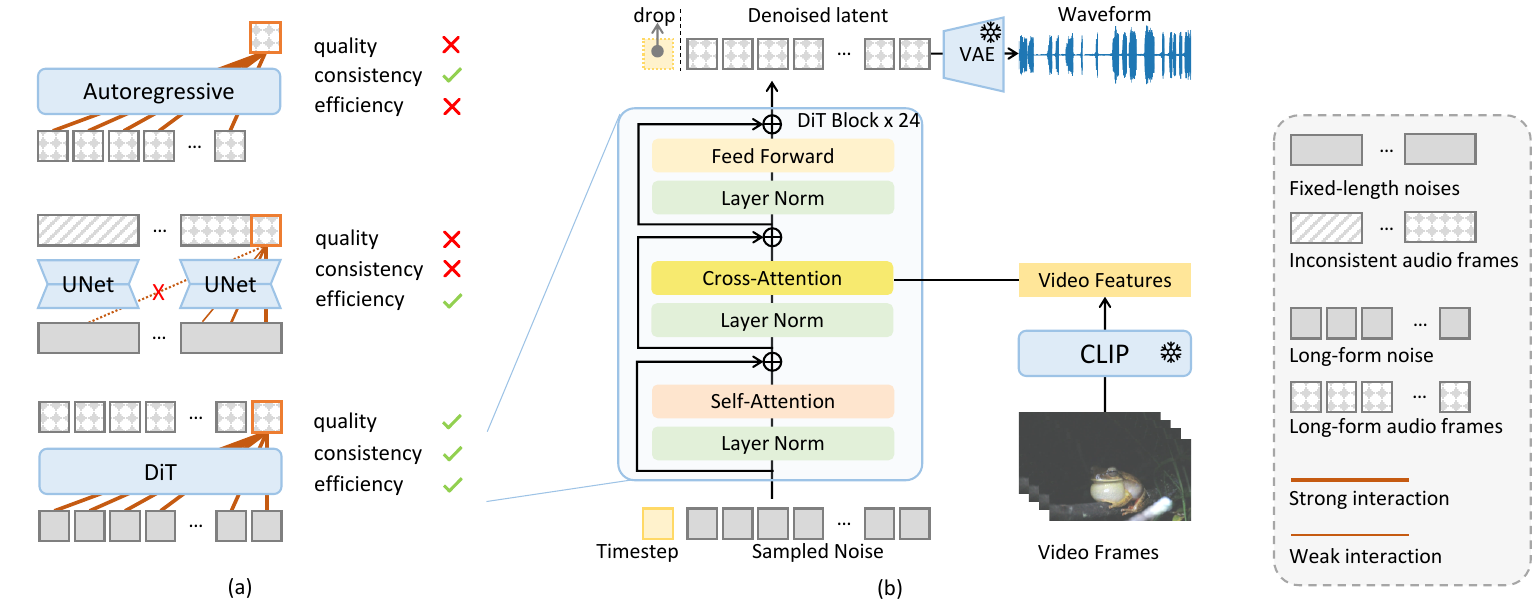}}
\caption{ \fontsize{9pt}{11pt} \selectfont 
(a) Comparison of three distinct long-form V2A methods. From top to bottom: autoregressive methods, UNet-based diffusions, DiT-based diffusions (our LoVA), characterized by inefficient one-by-one generation manner, inconsistent fixed-length splits generation, and our parallel processing of arbitrary-length sequences respectively. (b) Overview of LoVA. Capable of accepting videos of any length, it samples and denoises on the corresponding length of the latent noise sequence and then decodes it to generate audio of any length.
}
\vspace{-0.3cm}
\label{fig2}
\end{figure*}

When adapted to long-form videos, current autoregressive and UNet-based diffusion V2A models both exhibit limitations.
As depicted in Figure~\ref{fig2}(a):
(1) \textit{Autoregressive} methods model audio as a series of audio frames (i.e., tokens). Ideally, they can generate an infinite number of audio frames without truncation. 
However, this one-by-one generation process leads to low efficiency for long audio sequences.
It also yields lower audio quality compared to diffusion models due to frame discretization~\cite{wang2024tiva}.
(2) \textit{UNet-based}\cite{ronneberger2015u,rombach2022high} \textit{diffusion models}~ struggle with long-range relation modeling, with generation performance being constrained by the length of the training data~\cite{huang2023make}, a limitation confirmed by prior studies~\cite{chen2021transunet, zhou2021nnformer, hatamizadeh2022unetr} and our experimental results (Section.~\ref{comparison results}). 
To better accommodate long-form V2A, these models split long videos into shorter clips, equivalent to their pretraining data length, generate audio for each clip, and then concatenate these to form the final long audio.
However, such splitting process can result in inconsistencies, i.e., with distinct sounds from the same video. 
This is evident in Figure~\ref{case} with results from DiffFoley~\cite{luo2024diff}, TiVA~\cite{wang2024tiva}, and FoleyCrafter~\cite{zhang2024foleycrafter}, where short 8s/10s audio clips exhibit clear mel-spectrogram boundaries and structural differences,
thereby reducing the quality of the concatenated long audio.
Thus, balancing efficiency, consistency, and quality in long-form V2A remains a significant challenge for existing methods.


To address the aforementioned challenge, we introduce LoVA, a Long-form Video-to-Audio generation model that is designed to handle long-duration problem. 
As depicted in Figure~\ref{fig2}(a), the expected long-form V2A model should possess the capabilities of:
(1) maintaining the variable-length audio as a sequence of lossless frames to ensure quality; 
(2) modeling the full sequence interactions among frames, rather than the localized interactions learned by convolutional UNets, to ensure feasibility and consistency when extending to long sequences; 
(3) generating multiple frames in parallel for efficiency. 
Diffusion Transformer (DiT)\cite{peebles2023scalable} treats latent data as token sequences in the diffusion process, aligning well with the sequential nature of audio. It also demonstrated promising results in generating high-quality images\cite{peebles2023scalable}, videos~\cite{videoworldsimulators2024}, and audios~\cite{evans2024long} in an efficient parallel sequence generation manner.
Thus, we introduce DiT into the V2A domain and model the denoising process on noisy latent audio frames, termed as LoVA.
For long-form V2A problems, LoVA simply extracts extended video features and prepares correspondingly longer sequences of noisy audio frames for denoised lengthy audio generation, akin to a consistent frog croak over a 30s video as depicted in Figure~\ref{case}.

Furthermore, acknowledging the absence of research in the long-form V2A domain, we have established a long-form V2A evaluation based on a variable-length long video dataset UnAV100~\cite{geng2023dense}, as an addition to the current standard short-form evaluation. 
We conducted extensive experiments on this long-form evaluation to validate the performance of autoregressive, UNet-based diffusion, and DiT-based diffusion (our LoVA) methods, as well as their duration-extending characteristics in long-form V2A. 

Overall, our main contributions are as follows:
\begin{itemize}[leftmargin=*]

\item We first introduce the long-form generation problem in the V2A field and establish an evaluation framework for long-form V2A as a complement to existing V2A evaluations.
\item We first employ DiT into V2A area and propose a new model, LoVA, which is better suited for generating long-form audio than existing methods. 
\item We conducted extensive experiments on standard short-form and newly established long-form evaluation, validating the SOTA results achieved by LoVA. We also draw some duration-extending characteristics for different V2A methods. Demo samples for different V2A methods are available at https://ceaglex.github.io/LoVA.github.io/. 

\end{itemize}


\section{Method}

The long-form V2A task aims to generate an audio sequence $a$ of equivalent duration from any given long video $v$. 
We introduce LoVA, a Latent Diffusion Transformer designed for this task. 
As depicted in Figure~\ref{fig2}(b), LoVA preprocesses long-form video into features, applies denoising on a noise sequence of corresponding length, and eventually generates long-form audio through VAE decoding. 
We will sequentially elucidate the preliminary knowledge of Latent Diffusion Model (LDM)~\cite{rombach2022high}, the Architecture of LoVA and its training in the following subsections.

\subsection{Preliminary: V2A LDMs}



Given audio-video pairs $(a,v)$, the typical V2A LDM compresses $a$ into latent variables $z$ (i.e.,$z_0$ ) using a VAE encoder , and encodes $v$ into conditional video features $c$. 
A diffusion process then introduces Gaussian noise $\epsilon$ to the clean latent $z_0$ based on timestep $t$ and predefined noise schedule $\bar{\alpha_1}, ..., \bar{\alpha_t}, ..., \bar{\alpha_T}$: $z_t = \sqrt{\bar{\alpha_t}}z_0 + \sqrt{1-\bar{\alpha_t}}\epsilon, \epsilon \sim N(0,1)$.

During the denoising process, LDM aims to recover $z_0$ from $z_T$ by progressively estimating the added noise at each timestep $t$, given the condition $c$ and input noisy data $z_t$:

\vspace{-2mm}

\begin{equation}
\hat{\epsilon}_t = D(z_t, c, t).
\label{eq:predict}
\end{equation}

\vspace{-1mm}

The training objective is to minimize the L2 loss between the added noise \(\epsilon\) and the predicted noise \(\hat{\epsilon}_t\) at each step $t$:

\vspace{-1mm}

\begin{equation}
\mathcal{L} = ||\hat{\epsilon}_t - \epsilon||^2.
\label{eq:loss}
\end{equation}

\vspace{-1mm}

\subsection{Architecture of LoVA}


As shown in Figure~\ref{fig2}, LoVA has three components: an audio VAE $V$, a video encoder ${\rm CLIP}$ and a DiT-based denoiser $D$.

\noindent\textbf{(1) Audio VAE}:
LoVA employs a 1D-Conv-based VAE~\cite{evans2024stable, huang2023make} to compress the audio waveform $a \in [n, T]$, where $n$ and $T$ are the audio channels and time length. 
The resultant latent data is $z_0 = V(a) \in [n, T', h]$, with $T'$ and $h$ denoting the compressed time length and latent space size.



\noindent\textbf{(2) Video Encoder}:
Numerous previous works~\cite{sheffer2023hear, wang2024tiva} have demonstrated the effectiveness of CLIP~\cite{radford2021learning} in V2A task. 
For a video composed of a sequence of frames $v: [f_1, \ldots, f_i, \ldots, f_N]$, LoVA also uses the CLIP visual encoder to extract features from each video frame and concatenate them to form the video condition \(c\):

\vspace{-2mm}

\begin{equation}
\begin{aligned}
c_i &= {\rm CLIP} (f_i),
\\
c &= {\rm Concat}([c_1,  \ldots ,c_i, \ldots, c_N]) \in [N, h_C],
\end{aligned}
\end{equation}

\vspace{-1mm}
where \(N\) is the number of frames and \(h_C\) is the CLIP hidden size.
To further assist the model in learning the relationship between the audio sequence and video frames when extending to long-form V2A generation, we add a learnable positional embedding layer \(PE_c\) to the video condition \(c\):

\vspace{-1mm}

\begin{equation}
\begin{aligned}
p_c &= PE_c([1,...,i,...,N]) \in [N, h_C],
\\
c &= c + p_c.
\end{aligned}
\end{equation}

\vspace{-1mm}

Finally, the video frame features \(c\) are fed into the DiT denoiser as a visual condition to guide the denoising process.

\noindent\textbf{(3) DiT Denoiser}:
Diffusion Transformer (DiT)~\cite{peebles2023scalable} is a novel diffusion structure that integrates the denoising diffusion models~\cite{ho2020denoising} with the Transformer architecture~\cite{vaswani2017attention}. 
In LoVA's DiT denoiser, before being fed into each DiT block, the noised audio latent sequence \(z_t\) is first processed with an additional positional embedding layer \(PE_z\):
\vspace{-1mm}
\begin{equation}
\begin{aligned}
p_z &= PE_z([1,...,i,...,T']) \in [T', h],
\\
z_t &= z_t + p_z.
\end{aligned}
\end{equation}

\vspace{-1mm}
Timestep \(t\) is embeded and appended at the beginning of the input sequence. 
Conditional input \(c\) is processed by cross-attention layers in DiT block.
Finally, conditioned on timestep $t$ and video features $c$, DiT denoiser takes $z_t$ as input tokens to estimate noise at each timestep, as formalized in Equation~\ref{eq:predict}. 

\subsection{Training and Inference of LoVA}

In the optimization phase of LoVA, the Audio VAE and Video Encoder are maintained frozen, as per~\cite{evans2024stable,radford2021learning}. The DiT Denoiser, including all blocks, $PE_c$, $PE_z$, and time embeddings, undergoes training. The training is governed by the L2 Loss as described in Equation~\ref{eq:loss}. 
During inference, LoVA can accommodate videos of arbitrary lengths, handling variable-length video conditions and noisy latent sequences through the extension of $PE_c$ and $PE_z$. Finally, variable-length audio is obtained through VAE decoding.

\section{Experimental Settings}

\subsection{Implementation Details}
\label{Dataset and Implementation Details}

We implement a two-phase training approach: \textit{pre-training} with short-form data and then \textit{fine-tuning} with long-form data, referred to as \textbf{LoVA (w/o tuning)} and \textbf{LoVA (w/ tuning)} respectively. 
Throughout both phases, the weights of LoVA's Audio VAE and Video Encoder remain frozen~\cite{evans2024stable,radford2021learning}. 
For the pre-training phase, the DiT Denoiser, $PE_c$, $PE_z$, and time embedding undergo training. 
During the fine-tuning phase, updates are only applied to the embedding layers $PE_z$, $PE_c$, and the final DiT block. 
Beside, We sample our audio at 44.1kHz and video frames at 8 FPS. In the inference stage, we set the guidance scale to 5.0, and employ the DPM++ 3M SDE sampler~\cite{karras2022elucidating} to execute denoising over 150 steps.

\subsection{Datasets}

LoVA (w/o tuning) utilizes AudioSet-balanced~\cite{gemmeke2017audio} and VGGSound~\cite{chen2020vggsound} datasets, encompassing 20,280 and 180,379 10-second videos respectively. 
LoVA (w/ tuning) employs the UnAV100 dataset~\cite{geng2023dense}, made up of 6,489 videos ranging from 10 to 60 seconds. 
To assess LoVA's performance against baselines in short-form V2A generation, we use the VGGSound \cite{chen2020vggsound} test set of 15,273 10-second videos. 
For long-form V2A generation evaluation, we utilize the UnAV100~\cite{geng2023dense} test set, comprising 2,167 cases with an average duration of 42.1s.

\subsection{Baselines}
We implement the public code of five baselines to
replicate the results.: SpecVQGAN~\cite{SpecVQGAN_Iashin_2021}, IM2WAV~\cite{sheffer2023hear}, DiffFoley~\cite{luo2024diff}, TiVA~\cite{wang2024tiva}, and FoleyCrafter~\cite{zhang2024foleycrafter}, in which the first two are auto-regressive models while the other three are diffusion-based models. 
To ensure a fair comparison, we adapt all of them for long-form V2A generation.
For the autoregressive baseline SpecVQGAN, we use the long-form video as input, adjust the generated sequence length, and obtain aligned long-form audio. For the three diffusion-based baselines, we divide the original video into short-form fixed-length clips(8 seconds or 10 seconds consistent with their training settings), generate corresponding audio separately, and then concatenate the generated audio segments. It should be mentioned that for IM2WAV we use the same divide-generate-concatenate procedure due to its slow inference speed.

\subsection{Metrics}

Following previous works~\cite{SpecVQGAN_Iashin_2021, liu2023audioldm,liu2024audioldm,luo2024diff,sheffer2023hear}, we apply widely-used Fréchet Audio Distance (FAD)~\cite{kilgour2018fr}, Inception Score (IS)~\cite{salimans2016improved}, and mean KL-divergence (MKL) to evaluate the quality of generated audio. 
To ensure a fair comparison and eliminate the effect of different sampling rate, we downsample the generated audio from LoVA's to 16kHz and then resample them to the required sampling rate of classifiers (16kHz for VGGish~\cite{hershey2017cnn}, 32kHz for PaSST~\cite{koutini2021efficient} and PANN~\cite{kong2020panns}). 
Since these audio classifiers are trained on 10-second audio data, they cannot be directly applied to the evaluation of long-form audio. Thus for the evaluation of long-form V2A, we segment the generated audio into 10-second clips with 5-second overlapping windows. For FAD, we average features from all audio clips to get the final feature of long-form audio. For IS and MKL, following previous works~\cite{evans2024fast}, we get the mean results of classification logits and then apply a softmax. 

Besides, we introduce the number of inferences per audio (Num.Infer.) as an indicator of potential inconsistency. 
We also randomly select 40 videos from UnAV100 test set for human evaluation. Evaluators are asked to give a 5-level Likert scale on 4 aspects:Sound Quality (SoundQua.), Semantic Relevance (Sem.Rel.), Consistency (Cons.) and Overall quality (Overall).

\begin{table*}[t]
\caption{ \fontsize{9pt}{11pt} \selectfont  \MakeUppercase{Comparison of LoVA with baselines on VGGSound and UnAV100 benchmark.} We employ multiple classifiers to evaluate audio quality (``VGG'' denotes VGGish, ``PaSST'' denotes PaSST, and ``PANN'' denotes PANN). The best score on each metric is highlighted with bold type and the second best score is in underline. 
} 
\centering
\renewcommand\arraystretch{0.95} 


\scalebox{0.93}{
\begin{tabular}{l|c|ccccc|ccccc|cccc|c}

\toprule

\multicolumn{2}{c}{} & \multicolumn{5}{|c|}{VGGSound} & \multicolumn{10}{c}{UnAV100}  \\

\midrule

& Sampling & FAD$\downarrow$  &  KL$\downarrow$   &  KL$\downarrow$ &  IS$\uparrow$ &  IS$\uparrow$ & FAD$\downarrow$  &  KL$\downarrow$   &  KL$\downarrow$ &  IS$\uparrow$ &  IS$\uparrow$ &  Sound & Sem. & Cons. & Over  & Num.  \\
\textbf{Method}& Rate {\scriptsize{(kHz)}}&{\scriptsize{(VGG)}} &{\scriptsize{(PANN)}}&{\scriptsize{(PaSST)}} &{\scriptsize{(PANN)}} &{\scriptsize{(PaSST)}} &{\scriptsize{(VGG)}} &{\scriptsize{(PANN)}}&{\scriptsize{(PaSST)}} &{\scriptsize{(PANN)}} &{\scriptsize{(PaSST)}} & Qua. $\uparrow$ & \quad Rel. $\uparrow$ & $\uparrow$ & \quad all $\uparrow$ & \quad Infer.$\downarrow$ \\
       
\midrule
\multicolumn{13}{l}{\textcolor[RGB]{200,200,200}{\textit{AutoRegressive}}}\\
\midrule

SpecVQGAN   & 22.05 & 6.26 & 3.16 & 3.12 & 4.00  & 3.77 & 9.21 & 2.28  & 2.17 & 2.84  & 2.52 & -- & -- & -- & -- &\textbf{1.00} \\

IM2WAV      & 16    & 5.77 & 2.28 & 2.24 & 5.77 & 5.19 & 6.99 & 1.10 & \textbf{1.05} & 4.32 & 4.28 & 2.81 & \underline{3.12} & \underline{3.21}  & \underline{2.93} &  \underline{4.64} \\

\midrule
\multicolumn{13}{l}{\textcolor[RGB]{200,200,200}{\textit{Diffusion}}}\\
\midrule

DiffFoley   & 16    & 6.10 & 2.76  & 2.88 & 8.12 & 9.56 & 7.74 & 1.22  & 1.28 & 4.42 & 5.02 &  2.89  & 3.02  & 2.94 & 2.79  & 5.70  \\

FoleyCrafter& 16    & 2.34 & 2.29 & 2.28 & 8.53 & \underline{9.83} & \underline{2.82} & \underline{1.06} & \textbf{1.05} &  \underline{6.91} & \underline{7.47}  & 3.15 & 3.06  & 3.13  & \underline{2.93} & \underline{4.64}\\

TiVA        & 16    & \textbf{1.05} & \underline{2.13} & \textbf{2.00} & \underline{9.31} & 8.02& 6.36 & 1.48 & 1.54 & 3.11 & 2.77  & \textbf{3.50} & 2.78  & 2.79  &  2.76 & \underline{4.64} \\

\midrule

LoVA (w/o tuning)       & \textbf{44.1}  & \underline{1.70} & \textbf{2.06} & \underline{2.10} & \textbf{9.69} & \textbf{9.87} & \textbf{2.44} & \textbf{1.05} & \underline{1.06} & \textbf{7.69} & \textbf{7.96}   &  \underline{3.42}  & \textbf{3.55} & \textbf{3.81}  & \textbf{3.45} &\textbf{1.00}\\

LoVA (w/ tuning)       & \textbf{44.1}  & \underline{1.70} & \textbf{2.06} & \underline{2.10} & \textbf{9.73} & \textbf{9.91}   & \textbf{2.42} & \textbf{1.05} & \underline{1.06} & \textbf{7.71} & \textbf{7.96}  &  \underline{3.42} & \textbf{3.56} & \textbf{3.82} & \textbf{3.51} &\textbf{1.00}\\





\bottomrule
\end{tabular}
}
\label{tab1}

\vspace{-0.4cm}

\end{table*}


\section{Experimental Results}

\subsection{Comparison with SOTA models}




To evaluate the performance of LoVA on long-form V2A generation, we compare LoVA with baselines on the UnAV100 dataset. 
As shown in Table~\ref{tab1}, being trained on the same short-form data without any fine-tuning, LoVA (w/o tuning) outperforms all other baselines on 4 out of 5 autonomous metrics and 3 out of 4 human evaluation metrics, with the fewest Num.Infer. It proves the effectiveness of DiT model in long-form V2A tasks. 
Utilizing DiT, LoVA generates high sampling rate audio that is ~6 times longer than current UNet-based diffusion models. 
Besides, being fine-tuned on the long-form dataset, LoVA (w/ tuning) achieves the best performance regarding most metrics, considering both objective and subjective evaluations.
This remarkable performance underscores LoVA’s superiority in handling long-form V2A generation. 

To evaluate the performance of LoVA for the short-form V2A generation, we conduct experiments on the VGGSound test set.
As shown in Table~\ref{tab1}, LoVA achieves comparable or even better performance to existing state-of-the-art V2A models. Specifically, LoVA  achieves the best performance in PANN KL, PANN IS and PaSST IS scores, and the second-best score in FAD and PaSST KL scores. These results indicate that LoVA can generate realistic audio and accurately capture semantic information from the video. 
In summary, LoVA shows the best results in both short-form and long-form V2A generation across most evaluation metrics.

\subsection{Comparison between Different Diffusion Denoiser}

We conduct ablation studies to validate the effectiveness of DiT modules when performing long-form V2A generation. 
Being similar to LoVA, some UNet-based diffusion models, like FoleyCrafter~\cite{zhang2024foleycrafter}, can also generate long-form audio by expanding the shape of latent space. 
On the UnAV100 benchmark, we split the video into sub-videos with different splitting durations and generate sub-audio for each sub-video individually. Then we obtain the long-form audio by concatenating all generated sub-audios.  
FoleyCrafter is adapted to different splitting durations by resizing the shape of latent space. 

\label{comparison results}

\begin{figure}[htbp]
\centerline{\includegraphics[width=0.45\textwidth]{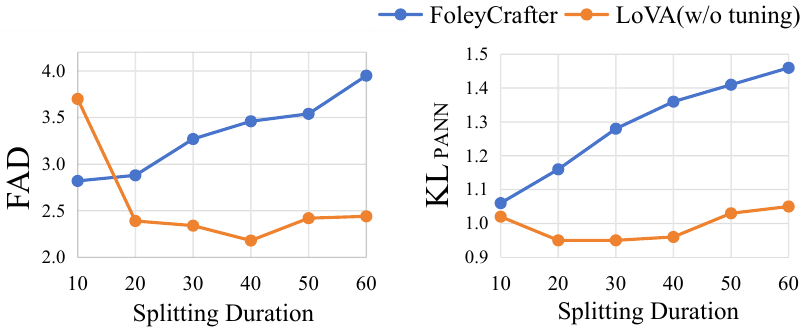}}
\caption{ \fontsize{9pt}{11pt} \selectfont Comparison of long-form audio generation ability between UNet and DiT structure. We choose FAD and KL to represent generated audios' quality. The experimennt is carried on UnAV100 test dataset. Different splitting durations mean different sub-videos' and generated sub-audios' durations per inference.}
\label{comp}
\vspace{-0.4cm}
\end{figure}


As shown in Fig~\ref{comp}, as the splitting duration increases, the FAD and KL metrics degenerate. The best scores are achieved when the splitting duration is 10 seconds, aligning with the training data's duration of FoleyCrafter. However, for DiT-based LoVA (w/o tuning), metrics do not exhibit obvious degeneration phenomenon as splitting duration increases. Notably, both FoleyCrafter and LoVA are trained on 10-second data only, yet perform differently on long-form audio generation. This difference highlights a critical limitation of the UNet structure when it extends to long-form audio, while proves DiT's effectiveness to handle long audio sequence.

\section{Conclusion and Future Work}

In this paper, we identify the significant gap between current V2A models and real-world V2A applications, particularly in generating long-form audio. To address this, we introduce a new task termed long-form video-to-audio generation.
Besides, 
we introduce LoVA, a DiT-based V2A generation model,
which is tailored for long-form V2A generation tasks.
Experimental results indicate that LoVA shows SOTA performance than previous models on both the 10-second VGGSound and long-form UnAV100 benchmarks, excelling in audio quality, sampling rate, and supported duration. In future work, we plan to (1) further explore temporal synchronization between video and audio, and (2) investigate methods for controlling generated audio, including text and duration control.

\section*{Acknowledgments}

This work is supported by the National Natural Science Foundation
of China (No.62276268) and ZHI-TECH GROUP.
\clearpage

\bibliographystyle{IEEEtran}
\bibliography{asset/VTA_variablelength}

\end{document}